\documentclass{an}
\usepackage{graphicx}
\usepackage{times}
\newcommand{\ba}{\begin{eqnarray}}
\newcommand{\ea}{\end{eqnarray}}

\newcommand{\Tgr}{T_{\tiny{\textrm{gr}}}}

\newcommand{\dotsgr}{\dot s_{\tiny{\textrm{gr}}}}
\newcommand{\pgr}{p_{\tiny{\textrm{gr}}}}
\newcommand{\qgr}{q^a_{\tiny{\textrm{gr}}}}
\newcommand{\Pigr}{\Pi^{ab}_{\tiny{\textrm{gr}}}}
\newcommand{\rhogr}{\rho_{\tiny{\textrm{gr}}}}
\newcommand{\ACal}{{\cal{A}}}

\newcommand{\JJ}{{\cal{J}}}

\newcommand{\HH}{{\cal{H}}}
\newcommand{\KK}{{\cal{K}}}
\newcommand{\RR}{{\cal{R}}^{(3)}}

\newcommand{\Del}{{\textrm{\bf{D}}}}

\newcommand{\dd}{{\rm{d}}}
\newcommand{\Da}{\delta^{(A)}}
\newcommand{\Drho}{\delta^{(\rho)}}

\newcommand{\Dmu}{\delta^{(\mu)}}
\newcommand{\Dmut}{\dot\delta^{(\mu)}}
\newcommand{\Dh}{\delta^{(\HH)}}

\newcommand{\Dhh}{\delta^{(h)}}
\newcommand{\Dhht}{\dot\delta^{(h)}}
\newcommand{\DKK}{\delta^{(\KK)}}
\newcommand{\Dk}{\delta^{(\kappa)}}
\newcommand{\Dig}{\Delta_0^{\tiny{\textrm{(+)}}}}
\newcommand{\Did}{\Delta_0^{\tiny{\textrm{(-)}}}}
\newcommand{\Jg}{\JJ^{\tiny{\textrm{(+)}}}}
\newcommand{\Jd}{\JJ^{\tiny{\textrm{(-)}}}}
\newcommand{\tbb}{t_{\textrm{\tiny{bb}}}}

\begin{document}

\Pagespan{789}{}
\Yearpublication{2014}%
\Yearsubmission{2014}%
\Month{11}%
\Volume{999}%
\Issue{88}%

\title{Gravitational entropy of cosmic expansion.}

\author{Roberto A. Sussman\fnmsep\thanks{Corresponding author:
  \email{sussman@nucleares.unam.mx}\newline}
}
\titlerunning{Gravitational entropy of cosmic expansion.}
\authorrunning{Roberto A. Sussman.}
\institute{
Instituto de Ciencias Nucleares, Universidad Nacional Aut\'onoma de M\'exico (ICN-UNAM). A. P. 70--543, 04510 M\'exico D. F.}

\received{04 Apr 2014}
\accepted{13 May 2014}
\publonline{2014 Aug 01}


\abstract{
We apply a recent proposal to define ``gravitational entropy'' to the expansion of cosmic voids within the framework of non--perturbative General Relativity. By considering CDM void configurations compatible with basic observational constraints, we show that this entropy grows from post--inflationary conditions towards a final asymptotic value in a late time fully non--linear regime described by the  Lema\^\i tre--Tolman--Bondi (LTB) dust models. A qualitatively analogous behavior occurs if we assume a positive cosmological constant consistent with a $\Lambda$--CDM background model. However, the $\Lambda$ term introduces a significant suppression of entropy growth with the terminal equilibrium value reached at a much faster rate.                
}

\maketitle

\section{Introduction}

The availability of a large amount of independent good quality precise observations has turned modern Cosmology into an exiting topic. Since the ``concordance'' or ``$\Lambda$--CDM'' model has been quite successful to provide an empiric fitting to these observations (\cite{paradigm1}; \cite{paradigm2}; \cite{planck}) and numerical n--body simulations provide a reasonably good description of our local Cosmography (\cite{nbody}), a great deal of research in Cosmology is based on linear perturbations on an FLRW ($\Lambda$--CDM) background (at scales comparable to the Hubble horizon) and Newtonian gravity for structure formation in sub--horizon scales. Since the $\Lambda$--CDM model does not explain the (yet unknown) fundamental nature of dark matter and dark energy, the currently dominant assumption in cosmological research is that undertaking these theoretical issues requires new early Universe physics: either quantum gravity or possibly new or modified gravity theories.  Nevertheless, there are still many open theoretical issues on the gravitational interaction at the cosmological scale that must be examined under the framework of  non--perturbative General Relativity (which, after all, is still our best ``classical''  gravity theory). 

One of the long standing open problems in General Relativity is the definition of a ``gravitational'' entropy providing a directionality to gravitational processes. From the old idea of the ``arrow of time'' (\cite{Penrose}; \cite{arrow}), research on this issue has produced two self--consistent proposals (\cite{CET};  \cite{HB}; \cite{susslar}) for a ``gravitational'' entropy that is different from (though possibly related with) the entropy of the field sources (hydrodynamical or non--collisional) or the holographic black hole entropies. 

We present in this article a summary of recently published research (\cite{susslar}) on the application of the  gravitational entropy proposal by \cite{CET} (to be denoted henceforth as the ``CET proposal'') to a cosmological context, and specifically to the expansion of cosmic voids (which dominate present day large scale CDM density distribution). For this purpose, we consider a non--perturbative framework 
\footnote{Gravitational entropy in a perturbative cosmological framework are examined by (Clifton et al. 2013) and by (\cite{Li}).}
through the class of exact spherically symmetric solutions of Einstein's equations with a dust source known as the Lema\^\i tre--Tolman--Bondi (LTB) models, as such models provide a simple but appropriate ``toy model'' description of cosmic voids (see comprehensive reviews in (\cite{kras}; \cite{book})).   

While density voids constructed with LTB models within the framework of non--perturbative General Relativity have been used as an alternative to the $\Lambda$--CDM paradigm (see review in (\cite{book})), it is important to remark that the usage of these models to examine the CET gravitational entropy is not (necessarily) in contradiction with this paradigm, as a nonzero $\Lambda$ term consistent with observations can easily be incorporated into their dynamics. Hence, we will examine the CET proposal for LTB models for the case $\Lambda=0$ and $\Lambda>0$.    

\section{Gravitational entropy.}   

The gravitational entropy defined in the CET proposal follows from an ``effective'' energy momentum tensor ${\cal T}^{ab}$ for the ``free'' gravitational field 
\footnote{The ``free'' gravitational field can be identified with the Weyl tensor. CET obtain the second order ``effective'' energy--momentum tensor ${\cal T}^{ab}$ through an irreducible algebraic decomposition of the Bell--Robinson tensor, the only fully symmetric divergence--free tensor that can be constructed from the Weyl tensor. See comprehensive discussion in (Clifton et al. 2013).}.
For Petrov type D spacetimes (``Coulomb--like'' fields), this tensor takes the form:
\ba \frac{{\cal T}^{ab}}{8\pi} = \rhogr u^au^b + \pgr h^{ab}+2q^{(a}_{\tiny{\textrm{grav}}} u^{b)}+\Pigr,\ea
with the  ``gravitational'' state variables $\rhogr,\,\pgr,\,\Pigr,\,\qgr$ (gravitational density, pressure, viscosity and heat flux) given by 
\ba 8\pi\rhogr &=& 2\alpha|\Psi_2|,\quad \pgr=\qgr=0,\nonumber\\ 8\pi\Pigr &=& \frac{\alpha|\Psi_2|}{2}(x^ax^b+y^ay^b-z^az^b+u^au^b).\label{effective}
\ea
where $\{u^a,x^a,y^a,z^a\}$ is an orthonormal tetrad, $\Psi_2$ is the conformal invariant for  of Petrov type D spacetimes and $\alpha$ is a constant to get the right units. Proceeding by analogy with entropy production in off--equilibrium hydrodynamical sources with 4--velocity $u^a$ in Eckart's frame (\cite{rund}), CET obtain the following Gibbs equation for the gravitational entropy growth:
\begin{equation}\Tgr\dotsgr = (\rhogr V)\dot{}=-V\sigma_{ab}\left[\Pigr+\frac{4\pi(\rho+p)}{3\alpha|\Psi_2|}E^{ab}\right],\label{gibbs}\end{equation}
where $V$ is a suitable local volume, $\sigma_{ab}$ is the shear tensor, $E^{ab}=u_au_b C^{acbd}$ is the electric Weyl tensor and the ``gravitational'' temperature $\Tgr$ is given by
\begin{equation} \Tgr = \frac{\left|\dot u_az^a+\HH+\sigma_{ab}z^az^b\right|}{2\pi},\label{Tgr}  \end{equation}
where $\dot u_a =u^b\nabla_a u_b$ is the 4--acceleration and $\HH\equiv h_c^b\nabla_b u^c/3$ is the isotropic Hubble expansion scalar. As commented by CET, the terms inside the brackets in the right hand side of (\ref{gibbs}) play the role of ``effective'' relativistic dissipation terms in the analogy with dissipative matter sources, though the actual sources are conserved, and thus the Gibbs equation (\ref{gibbs}) does not imply that they exchange energy or momentum with the free gravitational fields associated with (\ref{effective}). On the other hand, CET justify $\Tgr$ in (\ref{Tgr}) as a local ``gravitational''  temperature that reduces in semi--classical Unruh and Hawking temperatures in the appropriate limits (see further detail in (Clifton et al. 2013)).

Notice that FLRW models define a global ``gravitational'' equilibrium state characterized by $\dotsgr=0$ holding everywhere (for all $t$ and all fundamental observers), as for these models we have $\sigma_{ab}=\dot u_a=0$, while $\Tgr=|\HH|/(2\pi)>0$. As a consequence, the notion of gravitational entropy is intimately linked to the deviation from homogeneity inherent in the gravitational interaction. However, not every deviation from inhomogeneity can be associated with a physically plausible gravitational process. Hence, the CET gravitational entropy proposal must be tested through the fulfillment of the condition for entropy growth  
\begin{equation} \dotsgr \geq 0,\label{dotsgr}\end{equation}
that follows from (\ref{gibbs}) and (\ref{Tgr}), which should provide a directionality to gravitational processes when implemented in actual solutions of Einstein's equations. 

\section{LTB dust models.}

In order to probe the CET proposal, and specifically the entropy growth condition (\ref{dotsgr}) on LTB dust models, we describe the latter by the following FLRW--like metric element:           
\ba  \dd s^2 = \dd t^2+ a^2\left[\frac{\Gamma^2\dd r^2}{1-\KK_{q0}r^2}+r^2\dd\Omega^2\right],\label{ltb2}\ea  
where $\dd\Omega^2=\dd\vartheta^2+\sin^2\vartheta\dd\phi^2$,\,  $a=a(t,r)$ and $\Gamma =1+ra'/a$ with $a'=\partial a/\partial r$, while $K_{q0}=K_q(t_0,r)$ is defined in equation (\ref{KK}) further ahead (the subindex ${}_0$ will denote henceforth evaluation at present day cosmic time $t=t_0$, we remark that $a_0=\Gamma_0=1$). The main covariant objects of the models can be given as exact perturbations and fluctuations with respect to the q--scalars $\rho_q,\,\KK_q,\,\HH_q$
\footnote{These q--scalars can be related to a weighted scalar average of the covariant scalars $\rho,\,\KK,\,\HH$. They are covariant LTB scalars that satisfy identical evolution equations as their analogous FLRW scalars, hence they define a domain dependent FLRW background and allow to characterize LTB models as exact perturbations. See comprehensive discussion on their properties in Sussman (2010a, 2010b, 2013a, 2013b, 2013c). }
\ba \rho &=&\rho_q\,[1+\Drho_q],\qquad \rho_q=\frac{\rho_{q0}}{a^3},\label{rho}\\
  \KK &=& \RR_q\,[1+\DKK_q],\qquad \KK_q=\frac{\KK_{q0}}{a^2},\label{KK}\\
  \HH &=& \HH_q\,[1+\Dh_q],\quad \HH_q =\frac{\dot a}{a},\label{HH}\\
  \sigma_{ab} &=& \Sigma\, \hbox{\bf{e}}_{ab},\qquad \Sigma =-\Del_q(\HH),\label{shear}\\
  E_{ab} &=&  \Psi_2\,\hbox{\bf{e}}_{ab},\quad \Psi_2=\frac{4\pi}{3}\Del_q(\rho),\label{EWeyl}
  \ea
where $\KK=\RR/6$ (with $\RR$ the Ricci scalar of hypersurfaces orthogonal to $u^a$),\, $\hbox{\bf{e}}_{ab}=h_{ab}-3n_an_b$ with $n_a=\sqrt{g_{rr}}\delta_a^r$, while the perturbations and fluctuations ($\Da_q$ and $\Del_q(A)$ for $A=\rho,\,\KK,\,\HH$) are defined as
\ba \Del_q(A)=A-A_q=\frac{r A'_q}{3\Gamma},\qquad \Da_q =\frac{\Del_q(A)}{A_q}\label{Dadef}.\ea
If $\Lambda=0$ (we look at the case $\Lambda>0$ in section 6), we can obtain the following closed analytic forms for the exact perturbations $\Drho_q,\,\Dh_q$ 
\ba \Drho_q &=& \frac{\Jg+\Jd}{1-\Jg-\Jd},\label{Drho}\\
\Dh_q &=& \frac{(2+\Omega_q)(\Jg+\Jd)-2(1-\Omega_q)\Dig}{6(1-\Jg-\Jd)},\label{Dh} 
\ea
where the q--scalar $\Omega_q = \Omega_q^{(m)}$ is defined by
\ba \Omega_q =\frac{8\pi\rho_q}{3\HH_q^2}=\frac{\Omega_{q0}}{\Omega_{q0}+(1-\Omega_{q0})\,a},\ea
and $\Jg,\,\Jd$ are the exact generalizations of the density growing and decaying modes of linear perturbation theory (\cite{sussmodes}):
\ba \Jg &=& 3\Dig\,\left[\HH_q(t-\tbb)-\frac{2}{3}\right],\label{Jg}\\
\Jd &=&  3\Did\,\HH_q,\quad \label{Jd}\\
\Dig &=& \frac{\Drho_0-\frac{3}{2}\DKK_0}{1+\Drho_{q0}},\quad \Did =\frac{r\,\tbb'}{3(1+\Drho_{q0})},\label{ampl}\ea
where $\tbb=\tbb( r)$ is the Big Bang time such that $a=0$ as $t=\tbb$ for all $r$ and $\HH_q(t-\tbb)=Y_q(\Omega_q)$ holds with $Y_q$ given by
\ba
Y_q = \frac{\varepsilon_0}{1-\Omega_q}\left[1-\frac{\Omega_q}{2\sqrt{|1-\Omega_q|}}\ACal\left(\frac{2}{\Omega_q}-1\right)\right],\ea
with $\varepsilon_0=1,\,\ACal=$ arccosh for ever--expanding hyperbolic models with negative spatial curvature ($0<\Omega_q<1$) and $\varepsilon_0=-1,\,\ACal=$ arccos for elliptic collapsing models with positive spatial curvature ($\Omega_q>1$).     

\section{Density voids in regular LTB models.}

Density void profiles are a generic feature in regular hyperbolic LTB models with $\Lambda=0$ (\cite{RadProfs}; \cite{sussmodes}). The conditions for density void profiles follow from the time asymptotic forms of (\ref{Drho}) for such models (\cite{susslar}):
\ba \Drho_q &\approx& -1 \quad\hbox{as}\quad t\to\tbb \quad (\Did\ne 0),\label{asympt1a}\\
    \Drho_q &\approx& O(a)\to 0 \quad \hbox{as}\quad t\to\tbb \quad (\Did= 0),\label{asympt1b}\\
    \Drho_q &\approx& \frac{\Dig}{1-\Dig}\quad \hbox{as}\quad t\to\infty.\label{asympt2}
\ea
Since $1-\Dig>0$ and $1+\Drho_{q0}\geq 0$ must hold if we demand absence of shell crossings (\cite{sussmodes}; \cite{susslar}), and considering the sign relation between $\Drho_q$ and $\rho'_q$ from (\ref{Dadef}), the condition for a time asymptotic void profile ($\Drho_q>0$ so that $\rho'_q>0$ as $t\to\infty$) follows from (\ref{asympt2}) and is simply a positive growing mode $\Jg>0$, which (from (\ref{Jg}) and (\ref{ampl})) implies: $\Dig>0$, since $2/3<Y_q<1$ holds everywhere for hyperbolic models.  If there is an asymptotic void profile ($\Dig>0$) and the decaying mode is nonzero ($\Did\ne 0$), then  an ``inversion'' of the density profile necessarily occurs: an initial over--density at $t\approx \tbb$ (because of (\ref{asympt1a})) evolves into a density void as $t\to\infty$ (because of (\ref{asympt2})). On the other hand, if the decaying mode is suppressed ($\Did=0$) and $\Dig>0$, then (\ref{asympt1b}) and (\ref{asympt2}) imply that the density void profile occurs for the whole time evolution (\cite{RadProfs}). 

\section{Entropy growth in LTB voids.}     

The ``gravitational'' state variables $\rhogr$ and $\Tgr$ in (\ref{effective}) and (\ref{Tgr}) take the following forms for LTB models
\ba
 8\pi\rhogr &=& =2\alpha|\Psi_2|=\frac{8\pi\alpha}{3}|\Del_q(\rho)|,\label{rhograv2}\\
  \Tgr &=& \frac{|\HH_q|\,|\, 1+3\Dh_q\,|}{2\pi}.\label{Tgrav2}
\ea
Since the local volume defined by $\HH =\dot\ell/\ell$ is given by $V=\ell^3 = a^3\Gamma$, the condition for entropy growth (\ref{dotsgr}) from (\ref{gibbs}) becomes
\begin{equation}
\dotsgr= \frac{2\pi\alpha\rho_{q0}}{3}\frac{\partial_{t}\left(\Gamma\left|\Drho\right|\right)}{|\HH_{q}||1+3\Dh_q|},\label{CETc12}
\end{equation}
which, considering that $\dot \Gamma =3\Gamma \Del_q(\HH)=3\Gamma\HH_q\Dh$ (from (\ref{shear})) and assuming that $\Gamma>0$ holds to avoid shell crossing singularities (\cite{RadProfs}; \cite{sussmodes}), yields after a long algebraic manipulation (see detail in (\cite{susslar})) the necessary and sufficient condition for entropy growth expressed in terms of a negative correlation between fluctuations of the energy density and the Hubble scalar:
\begin{equation} \dotsgr\geq 0 \quad \Leftrightarrow \quad \Del_q(\rho)\Del_q(\HH)\leq 0.\label{CETcond1}\end{equation}
For ever--expanding density voids in hyperbolic models we have $\rho_q\geq 0,\,\HH_q\geq 0$ everywhere, hence (\ref{CETcond1}) becomes
\begin{equation} \dotsgr\geq 0 \quad \Leftrightarrow \quad \Drho_q\Dh_q\leq 0.\label{CETcond2}\end{equation}
\begin{figure}
\begin{center}
\includegraphics[scale=0.3]{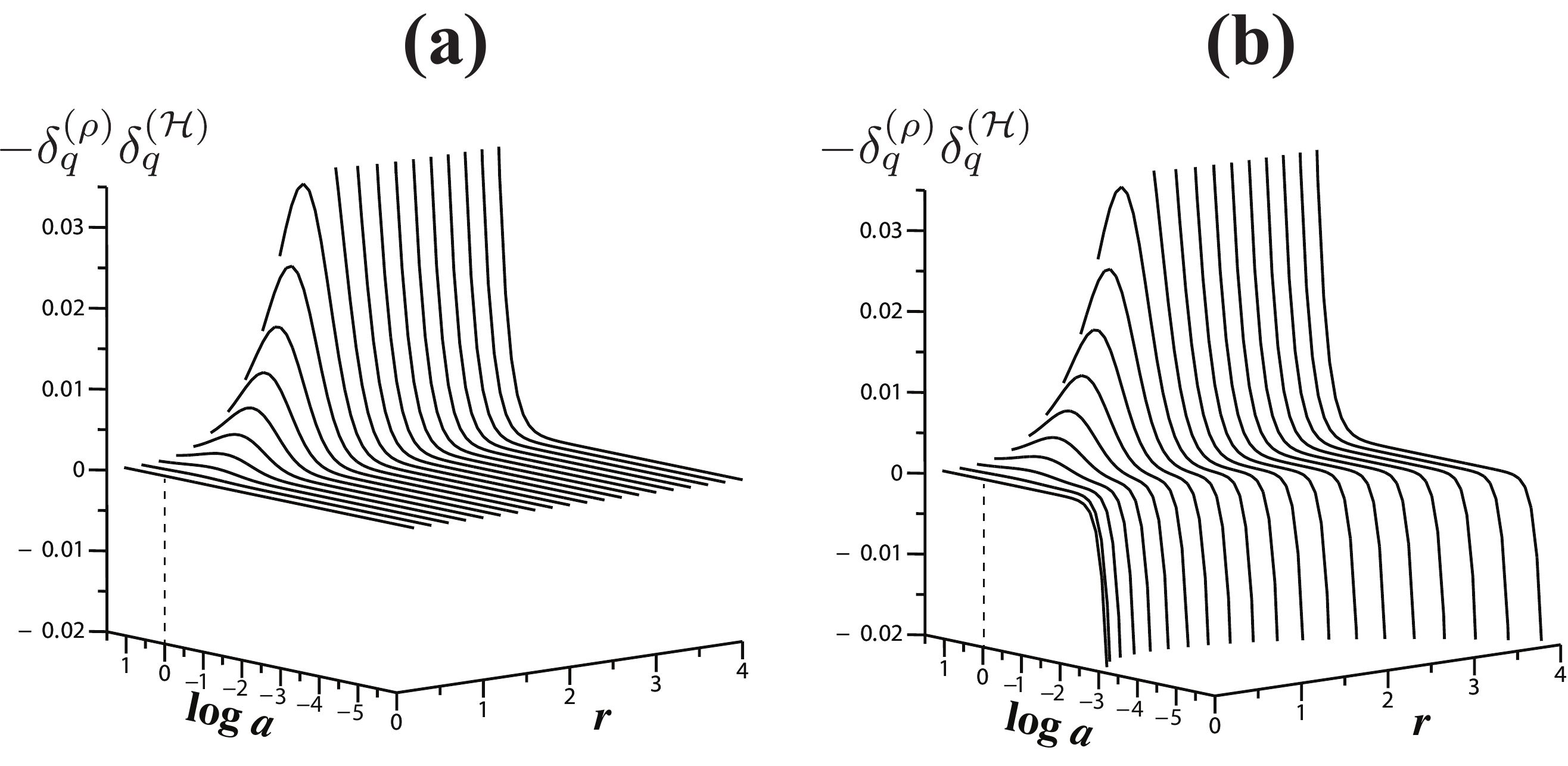}
\caption{{\bf Entropy production for LTB void models with zero and non-zero decaying mode.} The figure depicts $-\Drho_q\Dh_q \propto \dotsgr$ as a function of $\log a$ and $r$ marking the present day time as $a=a_0=1$ (notice that $a$ is a monotonic function of $t$ for hyperbolic models). Panel (a) corresponds to a void model with a suppressed decaying mode that is asymptotic to an Einstein de Sitter background and fits supernovae and CMB observations (taken from (\cite{February})). Panel (b) depicts a closely related model with a non-zero decaying mode (see (\cite{susslar})). Notice that entropy grows for large times in both cases, with the decrease due to the nonzero decaying mode ($\dotsgr<0$) taking place only for very early times $a< 10^{-3}$ in panel (b). }
\label{fig1}
\end{center}
\end{figure} 
Irrespective of whether the decaying mode is zero or nonzero, we have for late times $\dotsgr\to 0$ with $\dotsgr\geq 0$:
\ba \Drho_q\Dh_q &\approx& \left(\Dig\right)^2\Omega_q\left[1+\ln\left(\frac{\sqrt{\Omega_q}}{2}\right)\right]\leq 0\nonumber\\
 &\Rightarrow& \quad \dotsgr \geq 0,\ea
since $\Omega_q\ll 1$ holds as $t\to\infty$, and thus the logarithmic term inside the square brackets necessarily takes large negative values.  As depicted by figure 1, which compares the evolution of $\dotsgr$ for void models with zero and nonzero (but subdominant) decaying mode, the late time behavior $\dotsgr\to 0$ is the same regardless of whether the decaying mode is suppressed or not. Moreover, (\ref{CETcond2}) also holds irrespective of the sign of $\Dig$, which is an important result: entropy grows in the asymptotic time range of all hyperbolic models, whether the terminal density profile is that of a void ($\Dig\geq 0$) or an over--density ($\Dig\leq 0$) (\cite{sussmodes}). 

For early times the growth of entropy depends on the decaying mode (see the different early time behavior of the plots in panels (a) and (b) in figure 1). If the decaying mode is not suppressed ($\Did\ne 0$) we have $\Dh_q\approx \Drho_q/2\approx -1/2$ for $t\approx\tbb$ if $\Did\ne 0$, hence (\ref{asympt1a}) implies that $\dotsgr<0$ necessarily holds for these early times (near a non--simultaneous Big Bang). On the other hand, if the decaying mode is suppressed ($\Did=0$, simultaneous Big Bang), we obtain the opposite result: 
\begin{equation} \Drho_q\approx -\frac{2}{5}\Dig(\Omega_q-1)\to 0,\qquad \Dh_q\approx -\frac{1}{3}\Drho_q\to 0,\label{DrhoDh0}\end{equation}
which implies that (\ref{CETcond2}) is fulfilled as $t\to\tbb$. In fact, if $\Did=0$ then $\dotsgr\geq 0$ holds throughout the full time evolution of the hyperbolic models. While this result seems to suggest that models with a non-zero decaying mode should be discarded, this may be an excessively strong and unnecessary restriction, as an early time decrease of gravitational entropy from the decaying mode could simply be a signal that the dust source of LTB models no longer provides a physically viable description of cosmic dynamics in radiation dominated early times.
\begin{figure}
\begin{center}
\includegraphics[scale=0.45]{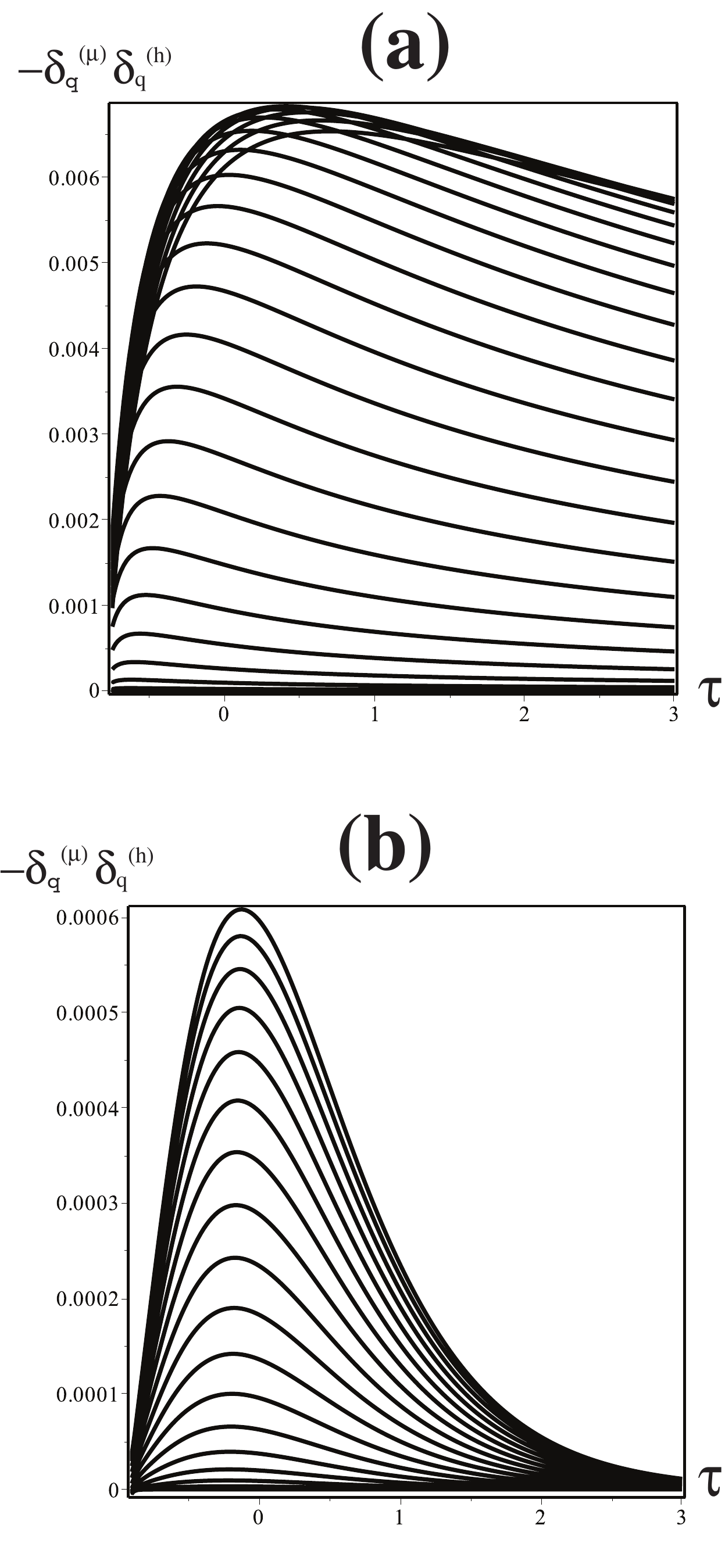}
\caption{{\bf Entropy production for LTB void models with zero and non-zero cosmological constant.} Both panels display the evolution of $-\Dmu_q \Dhh_q$, which is proportional to $\dotsgr$, as a function of the dimensionless time $\tau=H_{b0}(t-t_0)$, so that present cosmic time is marked by $\tau=0$ and Big Bang time is given by $\tau(\tbb)=-H_{b0} t_0$.  Panel (a) depicts a cosmic scale void in an open FLRW background with $\Omega_{b0}^{(\mu)}=0.3,\,\ \Omega_{b0}^{(\lambda)}=0$ and central value $\Omega^{(\mu)}_{c0}=0.08$. Panel (b) displays a similar void configuration in a $\Lambda$--CDM background $\Omega_{b0}^{(\mu)}=0.3,\,\ \Omega_{b0}^{(\lambda)}=0.7$ and central value $\Omega^{(\mu)}_{c0}=0.1$. It is noticeable that the $\Lambda$ term (panel (b)) suppresses the growth of gravitational entropy and forces a much faster limit $\dotsgr\to 0$. We do not claim that the displayed configurations are ``realistic'' nor that they fit supernovae or CMB data, as they have been conceived merely to illustrate the behavior of $\dotsgr$ while complying only with age constraints and observed values of the local Hubble constant (\cite{planck}).}
\label{fig1}
\end{center}
\end{figure} 
\section{The effect of $\Lambda>0$.}

If we assume $\Lambda>0$ in order to comply with the concordance observational paradigm, the condition for entropy growth (\ref{CETcond2}) remains valid for ever--expanding models ($\HH_q>0$), but there are no analytic closed forms for the perturbations $\Drho_q$ and $\Dh_q$. These perturbations can be computed numerically from solving the following evolution equations (\cite{sussDS2})
\ba \dot\mu_q &=&  -3\mu\,h_q,\label{ev1}\\
\dot h_q &=& -h_q^2 -\mu_q +\lambda,\label{ev2}\\
\Dmut_q &=& -3(1+\Dmu_q)\,h_q\Dhh_q,\label{ev3}\\
\Dhht_q &=& -(1+3\Dhh)\,h_q\Dhh_q +\nonumber\\
 &+& \frac{\mu_q(\Dhh_q-\Dmu_q)+\lambda\,\Dhh_q}{h_q},\label{ev4}
\ea 
subjected to the algebraic constraints
\ba h_q^2 &=& 2\mu_q -\kappa_q+\lambda,\label{c1}\\
2\Dhh_q &=& \Omega_q^{(\mu)}\Dmu_q +(1-\Omega_q^{(\mu)}-\Omega_q^{(\lambda)})\Dk,\label{c2}\ea
where $h_q=\HH_q/H_{b0}$,\,$\mu_q=4\pi\rho_q/(3H_{b0}^2)$,\,$\kappa_q=\KK_q/H_{b0}^2$ and $\lambda=8\pi\Lambda/(3H_{b0}^2)$, with $H_{b0}$ being a suitable FLRW background Hubble constant (not the observed Hubble constant), while the q--scalars associated with the Omega parameters for CDM and $\Lambda$  are 
\ba
\Omega_q^{(\mu)}=\frac{2\mu_q}{3h_q^2},\quad \Omega_q^{(\lambda)}= \frac{\lambda}{h_q^2}.\label{Om}\ea 
so that the background Omega factors $\Omega_{0b}^{(\mu)}$ and $\Omega_{0b}^{(\lambda)}$ are obtained as the limits of $\Omega_q^{(\mu)}$ and $\Omega_q^{(\lambda)}$ as $x=r/\sigma_0 \to \infty$, where $\sigma_0$ is an arbitrary adjustable length scale.  

We now use (\ref{ev1})--(\ref{Om}) to look at the time evolution of $\dotsgr$ through the product of perturbations in (\ref{CETcond2}). 
 The results are displayed in figure 2 for two cosmic voids complying with current data on the local Hubble constant $H_0 \approx 67 \,\,\hbox{km}/(\hbox{sec Mpc})$ and cosmic age constraints $t_0 \approx 13.8$ Gys (\cite{planck}). For a present day CDM density profile given by
\begin{equation} 2\mu_q = \Omega_{b0}^{(\mu)}+\frac{\Omega_{c0}^{(\mu)}-\Omega_{b0}^{(\mu)}}{1+x^2}\end{equation}
where $\Omega_{c0}^{(\mu)}=\Omega_{q0}^{(\mu)}(0)$,  we assume for one of the voids (panel (a)) $\lambda=\Omega_{b0}^{(\lambda)}=0$ and an open FLRW dust background with $\Omega_{b0}^{(\mu)}=0.3$, while for the second void (panel (b)) we consider a $\Lambda$--CDM background with $\Omega_{b0}^{(\mu)}=0.3$ and $\Omega_{b0}^{(\lambda)}=0.7$ (we assume for both examples a simultaneous Big Bang, which implies from (\ref{Jd}) and (\ref{ampl}) a suppressed decaying mode $\Did=0$).  It is evident from comparing the evolution of $\dotsgr$ in panels (a) and (b) of figure 2 that a nonzero cosmological constant (panel (b)) keeps a positive entropy growth but at the same time has an important time asymptotic suppression effect. 
 
\section{Conclusion}

We have applied the CET gravitational entropy proposal of Clifton et al (2013) to examine entropy growth in expanding cosmic voids that emerge from appropriate post--inflationary conditions, as we assumed a suppressed decaying mode (example (a) of figure 1 and both examples in figure 2) and in one case (example (b) of figure 1) a very subdominant decaying mode (see invariant definition of density modes of LTB models in (\cite{sussmodes})). As shown in (\cite{susslar}), the conditions for entropy growth hold as long as the growing mode is dominant over the decaying mode, even if the latter is not strictly zero. 

We have also examined the effect of a $\Lambda$ term on void models compatible with basic observational constraints. As we can see from figure 2, $\dotsgr$ is larger by an order of magnitude for the void with $\Lambda=0$ (panel (a)), and also the terminal entropy value associated with the convergence $\dotsgr\to 0$ for large cosmic times occurs at a much faster rate for the void with $\Lambda>0$ (panel (b)). However, this rapid convergence of the gravitational entropy does not occur in the present cosmic time ($\tau=0$ in figure 2) but at cosmic times about three times our cosmic age. Evidently, we have only examined very idealized spherical expanding cosmic voids, and thus further research is needed to probe the CET proposal (and the proposal of (Hosoya et al. 2004)) on more general spacetimes, such as Szekeres models (\cite{sussbol}), and on the process of structure formation and gravitational collapse. In particular, we aim at studying the growth of these gravitational entropies in the context of the formation of virialized stationary structures, which may provide a connection with theoretical work done on n--body numerical simulations (\cite{nbody}) and Newtonian self--gravitational systems (\cite{Newtonian}), as well as research on various proposals on non--extensive entropy definitions (\cite{Tsallis}). This research is currently under way and will be submitted for publication in the near future.


\acknowledgements
I acknowledge financial support from grant SEP--CONACYT 132132.

\newpage

%


\begin{thebibliography}{}
%
%
%


\bibitem[Allen, Evrard \& Matz 2008] {paradigm2}  Allen S.W., Evrard A.E. and Mantz A.B.: 2011 {\it Ann. Rev. Astron. Astrophys.} {\bf 49} 409--470 


\bibitem[Bolejko \& Stoeger 2013]{bolstoeg} Bolejko K. and Stoeger W.: 2013, \emph{Phys Rev} D, {\bf 88}, 063529.


\bibitem[Bolejko et al. 2009]{book} Bolejko K., Krasi\'nski A., Hellaby C. and C\'el\'erier M.N.: 2009, {\it Structures in the Universe by exact methods: formation, evolution, interactions} Cambridge University Press, Cambridge


\bibitem[Chissari \& Zaldariaga 2011]{nbody} Chissari N. E. and Zaldariaga M.: 2011,  \emph{Phys Rev} D, {\bf 83}, 123505 

\bibitem[Clifton, Ellis \& Tavakol 2013]{CET} 
Clifton T., Ellis G.F.R. and Tavakol R.: 2013 \emph{Class. Quantum Grav.} \textbf{30},
  125009.
  

\bibitem[February et al. 2010]{February} February S., Larena J. et al: 2010,  {\it Mon. Not. Roy. Astron. Soc. } {\bf 405} 2231 


\bibitem[Frieman, Turner \& Huterer 2008] {paradigm1} Frieman J., Turner M. and Huterer D.: 2008, {\it Ann. Rev. Astron. Astrophys.} {\bf 46} 385--432


\bibitem[Hosoya,  Buchert \& Morita 2004]{HB} Hosoya A.,  Buchert T. and Morita M.: 2004  \emph{Phys. Rev. Lett.} {\bf 92}, 141302-1.


\bibitem[Li et al. 2012]{Li} Li N., Buchert T., Hosoya A., Morita M. and Schwarz D.J.: 2012, \emph{Phys Rev} D, {\bf 86} 083539 


\bibitem[Maartens 1996]{rund} Maartens R.: 1996, ``Causal Thermodynamics in Relativity''. Lectures given at the Hanno Rund Workshop on Relativity and Thermodynamics, University of Natal, 1996  (e--print{\tt  arXiv:astro-ph/9609119v1}) 


\bibitem[Padmanabhan 1990; Binney \& Tremaine 1987, Saslaw 1985]{Newtonian} Padmanabhan T.: 1990 {\it Phys Rep}, {\bf 188}, 5;  Binney J. and Tremaine S.: 1987 \emph {Galactic Dynamics}, Princetopn University Press, Princeton, N.J.; Saslaw W.C.: 1985, \emph{Gravitational Physics of Stellar and Galactic Systems}, Cambridge University Press, Cambridge, U.K.


\bibitem[Planck Collaboration 2013]{planck} Planck collaboration: Ade P.A.R. et al.: 2013, ``Planck 2013 results. XVI. Cosmological parameters'', e--print {\tt arXiv:1303.5076v2 [astro-ph.CO]}. 

\bibitem[Penrose 1979]{Penrose} Penrose R., 1979 \emph{General Relativity, an Einstein Centenary Survey},  edited by by Hawking S. W. and Israel W., Cambridge University Press, Cambridge. 

\bibitem[Plebanski \& Krasinski, 2006]{kras} Plebanski J and Krasinski A.: 2006, \emph{An Introduction to General Relativity and Cosmology},  Cambridge University Press, Cambridge.

\bibitem[Sussman 2010a]{RadAs} Sussman R.A.: 2010, \emph{Gen Rel Grav}, {\bf{42}}, 2813--2864 (2010)

\bibitem[Sussman 2010b]{RadProfs} Sussman R.A.: 2010, \emph{Class. Quantum. Grav.}, \textbf{27}, 175001, (2010)

\bibitem[Sussman 2013a]{part1} Sussman R.A.: 2013,  \emph{Class. Quantum Grav.}, {\bf 30}, 065015

\bibitem[Sussman 2013b]{part2} Sussman R.A.: 2013,  \emph{Class. Quantum Grav.}, {\bf 30}, 065016

\bibitem[Sussman 2013c]{sussmodes} Sussman R.A.: 2013,  \emph{Class. Quantum Grav.}, {\bf 30}, 235001, (2013)


\bibitem[Sussman \& Izquierdo 2011]{sussDS2} Sussman R.A. and Izquierdo G.: 2011, {\it Class Quantum Grav.}  {\bf 28} 045006 


\bibitem[Sussman \& Bolejko 2012]{sussbol} Sussman R.A. and Bolejko K.: 2012,  \emph{Class. Quantum Grav.}, {\bf 29}, 065018.


\bibitem[Sussman \& Larena 2014]{susslar} Sussman R.A. and Larena J.: 2014, \emph{Class. Quantum Grav.} {\bf 31}, 075021


\bibitem[Tsallis 2009; Plastino \& Plastino 2003; Taruya \& Sakagami]{Tsallis} Tsallis C.: 2009, \emph{Introduction to Nonextensive Statistical Mechanics}, Springer Science+Business Media, Berlin, Germany; Plastino A.R. and Plastino A.: 2003, \emph{Phys Lett} A, {\bf 174}, 384, (2003); Taruya A. and Sakagami M.: 2003,  \emph{Phys Rev Lett}, {\bf 90}, 181101.


\bibitem[Wainwright 1984; Bonnor 1986; Bonnor 1987; Pelavas \& Lake 2000] {arrow} Wainwright J.: 1984, \emph{Gen Rel Grav}, {\bf{16}}, 657; Bonnor W. B.: 1986, \emph{Class. Quantum Grav.} {\bf 3}, 495; Bonnor W. B.: 1987, \emph{Phys Lett} A, {\bf 122}, 305;  Pelavas N. and Lake K.: 2000, \emph{Phys Rev} D, {\bf 62}, 044009.



%
\end{thebibliography}
\end{document}